\documentclass[twocolumn,preprintnumbers,pra,superscriptaddress,showkeys]{revtex4-2}

\usepackage{amsmath,amssymb,amsfonts,bbm,graphicx,float,times,psfrag}
\usepackage[pdftex]{color}
\usepackage{amsmath,bm}
\usepackage[colorlinks, linkcolor=blue, citecolor=blue,  breaklinks=true]{hyperref}
\usepackage{mathtools,amsfonts,mathptmx}
\usepackage[utf8]{inputenc}
\usepackage[T1]{fontenc}
\usepackage{xcolor}
\usepackage{braket}
\usepackage[titletoc,title]{appendix}
\usepackage[nameinlink,capitalize]{cleveref}

\begin{document}
\title{Parametrically enhancing sensor sensitivity at an exceptional point}
\author{P. Djorwé}
\email{djorwepp@gmail.com}
\affiliation{Department of Physics, Faculty of Science, 
University of Ngaoundere, P.O. Box 454, Ngaoundere, Cameroon}
\affiliation{Stellenbosch Institute for Advanced Study (STIAS), Wallenberg Research Centre at Stellenbosch University, Stellenbosch 7600, South Africa}
\author{M. Asjad}
\email{asjad_qau@yahoo.com}
\affiliation{Mathematics Department, Khalifa University of Science and Technology, 127788, Abu Dhabi, United Arab Emirates}
\author{Y. Pennec}
\affiliation{Institut d’Electronique, de Microélectronique et Nanotechnologie, UMR CNRS 8520 Université de Lille, Faculté des sciences et technologies, 59652 Villeneuve d’Ascq France}
\author{D. Dutykh}
\affiliation{Mathematics Department, Khalifa University of Science and Technology, 127788, Abu Dhabi, United Arab Emirates}
\affiliation{Causal Dynamics Pty Ltd, Perth, Australia}
\author{B. Djafari-Rouhani}
\affiliation{Institut d’Electronique, de Microélectronique et Nanotechnologie, UMR CNRS 8520 Université de Lille, Faculté des sciences et technologies, 59652 Villeneuve d’Ascq France}
 
\begin{abstract}
We propose a scheme to enhance the sensitivity of Non-Hermitian optomechanical mass-sensors. The benchmark system consists of two coupled optomechanical systems where the mechanical resonators are mechanically coupled. The optical cavities are driven either by a blue or red detuned laser to produce gain and loss, respectively. Moreover, the mechanical resonators are parametrically driven through the modulation of their spring constant. For a specific strength of the optical driving field and without parametric driving, the system features an Exceptional Point (EP). Any perturbation to the mechanical frequency (dissipation) induces a splitting (shifting) of the EP, which scales as the square root of the perturbation strength, resulting in a sensitivity-factor enhancement compared with conventional optomechanical sensors. The sensitivity enhancement induced by the shifting scenario is weak as compared to the one based on the splitting phenomenon. By switching on parametric driving, the sensitivity of both sensing schemes is greatly improved, yielding to a better performance of the sensor. We have also confirmed these results through an analysis of the output spectra and the transmissions of the optical cavities. In addition to enhancing EP sensitivity, our scheme also reveals nonlinear effects on sensing under splitting and shifting scenarii. This work sheds light on new mechanisms of enhancing the sensitivity of Non-Hermitian mass sensors, paving a way to improve sensors performance for better nanoparticles or pollutants detection, and for water treatment.
\end{abstract}
\maketitle
\date{\today}
\section{Introduction} \label{Intro}
Exceptional Points (EPs), non-Hermitian degeneracy, are known in physical systems for their interesting counter-intuitive features and intriguing effects. Among the fascinating properties of EPs, we can mention non-reciprocity \cite{Thomas_2016,Lau_2018}, topological transport phenomena \cite{Xu_2023,Xu.2023}, stopping light \cite{Goldzak_2018}, loss-induced suppression and revival of lasing, pump-induced lasing death and unidirectional invisibility (see \cite{Peng_2014} and references therein). Recently, EPs have been used to enhance sensor sensitivity and versatile physical systems were proposed for this purpose including optical/photonic \cite{Hodaei_2017,Chen_2017}, microwave \cite{Nada_2018}, plasmonics \cite{Park.2020,Jiang_2022}, and mechanics/acoustics \cite{Rosa_2021,Cai_2022}. Coupled optomechanical systems, which result from a coupling between an electromagnetic field and a mechanical object \cite{Djorwe_2013,Aspelmeyer.2014}, have also been used to engineer EPs owing to their ability to induce gain and losses depending on their sideband pumping mechanism. EPs engineering in optomechanics have led to a number of interesting phenomena such as mass sensing improvement \cite{Djorwe.2019,Tchounda_2023}, topological energy transfer \cite{Chen_2023,Xu_2016}, and nonlinearly induced collective phenomena \cite{Djorwe_2018,Djorwe.2020} among others.

Exceptional point sensing is an approach which provides a better sensitivity compared to conventional sensing schemes, where the perturbation induces a shift in the frequency. The resulting shift/signal is generally weak and needs a pre-amplification process in order to be detected and analyzed \cite{Zhu_2021}. For the non-Hermitian sensors, however, the signal is self-amplified owing to the abrupt singularity and the topological feature at the EP, leading to a giant sensitivity enhancement without a pre-amplification requirement of the detected signal. This new  sensing scheme has recently attracted attention as a means to enhance the responsiveness of sensors, and has started to be practically implemented. For instance, exceptional point sensing has been used to monitor internal physiological states via wireless interrogation techniques \cite{Chen_2019}, and for an efficient wireless power transfer \cite{Assaw.2017,Song_2021}. Besides these interesting sensing applications probed at EPs, the implementation of EP sensors is still in its infancy stage and it seems to rise up scepticism regarding noise amplification, which induces a degradation of sensor’s performance \cite{Anderson.2023}. This noise issue can be of a fundamental nature owing to the eigenbasis collapse or of a technical nature related to the amplification mechanisms used for the realization of EPs. This question has been a bottleneck of EP-sensors, however, appropriate techniques have started to be used to handle and suppress this noise issue \cite{Kononchuk_2022,Suntharalingam_2023,Tuxbury_2022}. Therefore, tremendous research activities are still going on to tackle and to get rid of this limitation \cite{Chen_2020,Liu_2021,Wu_2023,Khanbekyan_2023}, and this includes nonlinear phenomena \cite{Li_2023,Bai_2022}.

The current work introduces a new mechanism to enhance the sensitivity of EP-sensors, which is based on a mechanical parametric driving effect. Our proposal consists of two coupled optomechanical systems, where the mechanical resonators are mechanically coupled. The parametric driving comes from a nonlinear modulation of the spring constants of the mechanical resonators \cite{Lemonde.2016,Bothner.2020}. We investigated two sensing schemes. The first one is based on the splitting \cite{Djorwe.2019} of the EP, resulting from a perturbation of the mechanical resonance frequencies, and the second one is related to the shifting \cite{Wang_2019,Mao_2023} of the EP, due to a mismatch between the mechanical dissipations. When parametric driving is switched on, we found that \emph{i}) the sensitivity is greatly enhanced, \emph{ii}) the EP is shifted towards low frequencies, and \emph{iii}) the strength of the optical driving field corresponding to the EP increases. Moreover, it has been observed that the overall sensitivity enhancement in the shifting scenario is weak compared to the enhancement captured with the splitting scheme. Furthermore, the output spectra and transmissions of the optical cavities have been investigated under some system parameters. The effect of perturbations, the parametric driving impact and thermal noise on the EP were analyzed via the spectra. While the effect of the other parameters is confirmed as aforementioned, it has been found that thermal noise amplifies the output signal, leading to better sensitivity. This work shows how sensing based on the splitting of the EP outperforms sensors operating in a scenario where the EP is shifted. Furthermore, it also paves a way towards improvement of sensing through nonlinear effects in optomechanics \cite{Djorw.2013,Leijssen.2017,Djorwe_2019}. 

The rest of the work is organized as follows. In \cref{model} we describe the model and derive the related dynamical equations. The sensitivity schemes based on splitting and shifting of the EP are presented in \cref{Sen.EP}. The transmissions and output spectra of the optical cavities are analyzed in \cref{Noise}, while \cref{Con} concludes our work. 
\section{Results} \label{resl}

\subsection{Model and dynamical equations} \label{model}
\begin{figure*}[tbh]
\begin{center}
\resizebox{0.33\textwidth}{!}{
\includegraphics{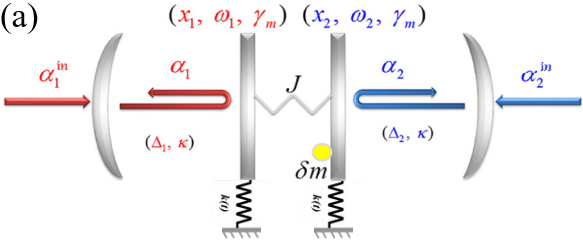}}
\resizebox{0.33\textwidth}{!}{
\includegraphics{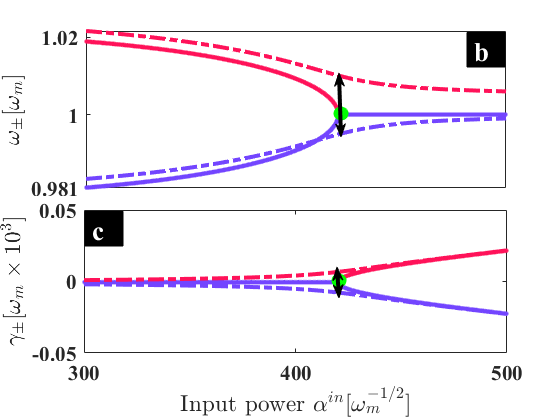}}
\resizebox{0.33\textwidth}{!}{
\includegraphics{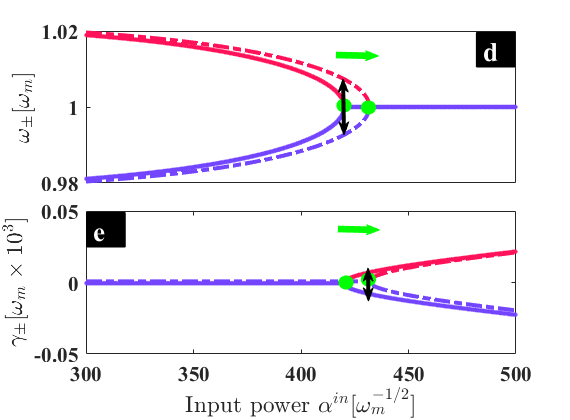}}
\end{center}
\caption{Overview of the schemes for $\chi=0$. (a) Sketch of our benchmark system. Cavity $1$ ($2$) is driven with red (blue) electromagnetic field to generate losses (gain). The spring constants of the involved mechanical resonators are modulated to induce parametric driving (see \cite{Lemonde.2016,Bothner.2020} for more details). Eigenmodes and the EP splitting (b,c) and  shifting (d,e) after a perturbation strength that induced a shift of $\delta\omega=5\times10^{-3}\omega_m$ in (b,c) (or $\delta\gamma=5\times10^{-3}\omega_m$ in (d,e)). The $\rm{EP}$ happens around the driving strength $\alpha^{\rm in}=420 \omega_m^{-1/2}$ for both sensors when there is no perturbation. The dimensionless experimentally feasible used parameters are \cite{Djorwe.2019}, $\kappa=0.1\omega_m$, $\Delta_1=-\omega_m$, $\Delta_2=\omega_m$, $\gamma_m=10^{-3}\omega_m$,  $g=2.5\times10^{-4}\omega_m$, and $J_m=2.2\times10^{-2}\omega_m$.}
\label{fig:Fig1}
\end{figure*}

We propose a two-cavity system with mechanically coupled vibrating mechanical resonators, as shown in \autoref{fig:Fig1}a. One cavity is driven by a red (blue) driving field that creates loss (gain) in the system. In addition, the mechanical resonators are parametrically driven for sensitivity enhancement prospect as it will be pointed out later on. The Hamiltonian of this system ($\hbar=1$) is given by the following equation:
\begin{equation}\label{eq:eq1}
 H =H_{\rm{0M}}+H_{\rm{PA}}+H_{\rm{int}}+H_{\rm{drive}}, 
 \end{equation}
where
\begin{eqnarray}  \label{eq:eq2}
H_{\rm{0M}}&:=&\sum_j \left(\omega_j b^{\dagger}_jb_j+\omega_c^{j}a^{\dagger}_ja_j -g_j a^{\dagger}_ja_j (b^{\dagger}_j+b_j) \right), \\
H_{\rm{PA}}&:=&\sum_j \frac{\chi_j}{2}(b^{\dagger 2}_je^{-{\rm i} \phi_d}+b^{2}_je^{{\rm i} \phi_d}), \\
H_{\rm{int}}&:=&-J_m(b_1b_2^{\dagger} + b_1^{\dagger}b_2),\\
H_{\rm{drive}}&:=&{\rm i} \sum_j ( {\rm E}_j  a^{\dagger}_j {\rm e}^{-{\rm i}\omega^j_{p} t} - \rm{E}^*_j   a_j {\rm e}^{{\rm i}\omega^j_{p} t} ).
\end{eqnarray}   
In \cref{eq:eq1}, $H_{\rm{0M}}$, $H_{\rm{PA}}$, $H_{\rm{int}}$, and  $H_{\rm{drive}}$ are the Hamiltonians describing the optomechanical cavities, the mechanical parametric amplification (\rm{MPA}), the interaction between the mechanical resonators and the optical driving fields, respectively. The annihilation (creation) operators related to the mechanical resonators and optical fields are $b_j$ ($b^{\dagger}_j$) and  $a_j$ ($a_j^{\dagger}$), respectively. The other parameters are the mechanical ($\omega_j$) and cavity ($\omega_c^j$) frequencies, the optomechanical coupling ($g_j$), the phonon hopping coupling rate ($J_m$) and the driving amplitude ($\rm{E_j}$). The parametric driving amplitude $\chi_j$ comes from the modulation of the spring constants of the mechanical resonators at frequency $2\omega_d$ and phase $\phi_d$  (see details in \cite{Lemonde.2016,Bothner.2020}). Throughout the work, we also assume $\phi_d=0$ and $\chi_1=\chi_2 \equiv \chi$. For the sake of simplicity and without loss of generality, we assume that the mechanical resonators are degenerated, and $g_1=g_2\equiv g$. We would like to mention that in the case of non-degenerated mechanical resonators, the mechanical parametric coefficient $\chi$ can be used to tune mechanical frequencies as it will be shown later on, leading to similar results. In the rotating wave approximation (RWA) and in the frame rotating at $\omega_p^j + \omega_d$, where $\omega_p^j$ is the $j^{\rm th}$ electromagnetic driving frequency, the above Hamiltonian yields: 
\begin{subequations}  \label{eq:eq3}
  \begin{align}
H_{\rm{0M}}&:=\sum_j \left(\tilde{\omega}_j b^{\dagger}_jb_j-\Delta_j a^{\dagger}_j a_j -g a^{\dagger}_j a_j (b^{\dagger}_j+b_j ) \right), \\
H_{\rm{PA}}&:=\frac{\chi}{2}\sum_j(b^{\dagger 2}_j+b^{2}_j), \\
H_{\rm{int}}&:=-J_m(b_1b_2^{\dagger} + b_1^{\dagger}b_2),\\
H_{\rm{drive}}&:={\rm i} \sum_j ({\rm E_j} a^{\dagger}_j - {\rm E_j}^* a_j),
    \end{align}       
\end{subequations}
where $\tilde{\omega}_j:=\omega_j - \omega_d$ and $\Delta_j:=\omega_p^j - \omega_c$, with the assumption that $\omega_c^1=\omega_c^2\equiv\omega_c$. 
In order to diagonalize the quadratic term in the above Hamiltonian,  we introduce the squeezing transformation $S(r_j)=\rm{exp}[r_j(b^2_j-b^{\dagger  2}_j)]$, where the squeezing parameter is defined as $r_j:=\frac{1}{4} \log{ \frac{\tilde{\omega}_j+\chi}{\omega_j-\chi }}$. By using the Bogoliubov transformation $b_{s_j}=b_j\cosh(r_j)+b^{\dagger}_j\sinh(r_j)$, and invoking the Heisenberg equation, one gets the following Quantum Langeving Equations (QLEs),
\begin{subequations}  \label{eq:eq4}
  \begin{align}
    \Dot{a_j} &=\left({\rm i}[\Delta_j + \tilde{g}(b_{s_j}+b^{\dagger}_{s_j})]-\frac{\kappa}{2}\right) a_j +  \sqrt{\kappa} \alpha^{\rm in}_j + \sqrt{\kappa} a^{\rm in}_j  , \\
    \Dot{b_{s_j}}&=-\left({\rm i}\Delta_m^j+\frac{\gamma_m}{2}\right)b_{s_j}+{\rm i} \tilde{J}_m b_{3-j}+{\rm i} \tilde{g}a^{\dagger}_j a_j +\sqrt{\gamma_m} \beta^{\rm in}_j.
 \end{align}       
\end{subequations}
In \cref{eq:eq4}, the driving amplitude $\rm{E}$ has been replaced by $\sqrt{\kappa}\alpha^{\rm in}$ with $\alpha^{\rm in}=\sqrt{\frac{P_{\rm in}}{\hbar \omega_p}}$, where  $P_{\rm in}$ is the input power. Also, the following tilded parameters have been defined,  $\tilde{g} =g {\rm e}^{-r}$, $\Delta_m^j=\sqrt{\tilde{\omega}_j^2-\chi^2}$, and  $\tilde{J}_m=J_m \cosh(2r)$. The optical and mechanical dissipation rates are captured by $\kappa_1=\kappa_2\equiv\kappa$ and $\gamma_1=\gamma_2\equiv \gamma_m$, respectively. Here  $a^{\rm in}_j$ and $\beta^{\rm in}_j$ are input noise operators of the ${\rm j}^{\rm th}$ cavity and mechanical modes with zero mean values.

The nonlinear set of equations displayed in \cref{eq:eq4} can be linearized by splitting the operators in terms of their mean values and some amount of quantum fluctuations as $a_j=\alpha_j+\delta \alpha_j$ and $b_{s_j}=\beta_j+\delta \beta_j$, where $\alpha_j=\langle a_j \rangle$ and $\beta_j=\langle b_{s_j} \rangle$. This well known procedure leads to classical mean field and fluctuations equations (see details in \cref{App.A}), where the detuning and optomechanical coupling are substituted with their effective quantities  $\tilde{\Delta}_j:=\Delta_j+2\tilde{g} {\rm Re}[\beta_j]$ and  $\tilde{G_j}=\tilde{g}\alpha_j$, respectively.  and we have assumed $\alpha_j$ to be a real number. 

In order to get an effective mechanical model of our system, we need to trace out the optical equations. For that purpose, we introduce the following slowly varying operators with tildes, $\delta\alpha_j=\delta\tilde{\alpha_j}{\rm e}^{{\rm i}\tilde{\Delta}_j t}$, $\delta\alpha^{\rm in}=\delta\tilde{\alpha^{\rm in}}{\rm e}^{{\rm i}\tilde{\Delta}_j t}$, and $\delta\beta_j=\delta\tilde{\beta_j}{\rm e}^{-{\rm i}\Delta_m^j t}$. By using these slowly varying operators, and by neglecting the terms oscillating at higher frequencies, the intracavity fields can be integrated out (see \cref{App.A}), and after some arrangements one obtains the following compact effective mechanical system:
\begin{equation} \label{eq:eq18}
    \Dot{\mathcal{O}}=\mathcal{M}\mathcal{O}+{\rm i}\sqrt{\Gamma} \mathcal{N}+\sqrt{\gamma_m}  \mathcal{O}^{\rm in},
\end{equation}
where $\mathcal{O}\equiv (\delta\beta_1, \delta\beta_2)^\top$, $\mathcal{N}\equiv (\delta\alpha^{\rm in}_1, \delta\alpha^{in\dagger}_2)^\top$, $\mathcal{O}^{\rm in}\equiv (\delta\beta^{\rm in}_1, \delta\beta^{\rm in}_2)^\top$, and the matrix
\begin{equation} \label{eq:eq19}
\mathcal{M}=
\begin{bmatrix}
-\left({\rm i}\Delta_m^1+\frac{\rm{\Gamma^1_{\rm eff}}}{2}\right) & {\rm i}\tilde{J}_m \\
{\rm i}\tilde{J}_m & -\left({\rm i}\Delta_m^2+\frac{\rm{\Gamma^2_{\rm eff}}}{2}\right) 
\end{bmatrix}.
\end{equation}
The effective damping are ${\rm \Gamma^{1,2}_{\rm eff}}=\gamma_m \pm  \rm{\Gamma}_{1,2}$, where the optically induced damping is $ {\rm \Gamma}_j:=\frac{4|\tilde{G_j}|^2}{\kappa}$, and from now on, we will assume that $ \rm{\Gamma}_1= \rm{\Gamma}_2 \equiv \rm{\Gamma}$.

\subsection{Sensitivity through a splitting and shifting of an exceptional point}\label{Sen.EP}
In order to get the eigenvalues of our effective mechanical system, we rewrite \cref{eq:eq18} in the Schr{\"o}dinger-like equation, ${\rm i}\partial_t \Psi =\rm{H_{\rm eff}}\Psi$ where $\Psi := (\delta\beta_1, \delta\beta_2)^\top$.  By ignoring noise terms, this leads to an effective Hamiltonian: 
\begin{equation} \label{eq:eq20}
\rm{H_{\rm eff}} = 
\begin{bmatrix}
\Delta_m^1-{\rm i}\frac{\rm{\Gamma^1_{\rm eff}}}{2} & -\tilde{J}_m \\
-\tilde{J}_m & \Delta_m^2-{\rm i}\frac{\rm{\Gamma^2_{\rm eff}}}{2} 
\end{bmatrix}.
\end{equation}

\begin{figure*}[tbh]
\begin{center}
\resizebox{0.45\textwidth}{!}{
\includegraphics{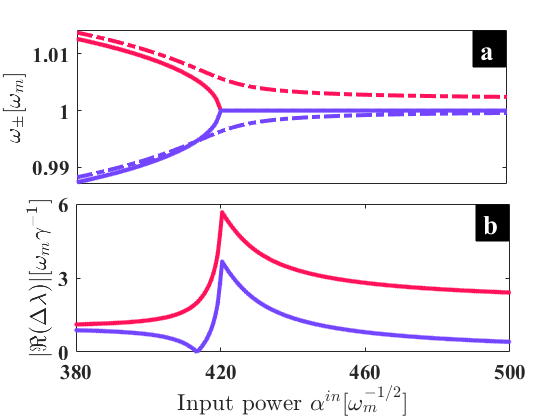}}
\resizebox{0.45\textwidth}{!}{
\includegraphics{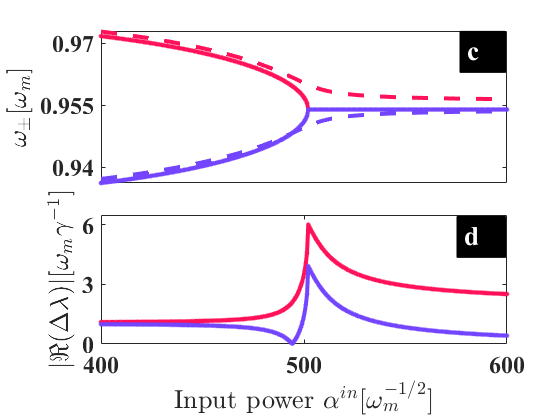}}
\end{center}
\caption{Parametric effect. Splitting of the eigenfrequencies for $\chi=0$ (f) and for $\chi=0.3\omega_m$ (h). The solid lines are before any perturbation and the dashed lines capture the perturbed behaviors resulting from a shift of $\delta\omega=2\times10^{-3}\omega_m$ in (a,b) and $\delta\gamma=2\times10^{-3}\omega_m$ in (c,d). Panels (b) and (d) are the induced splittings from (a) and (c), respectively.  The other used parameters are as in \autoref{fig:Fig1}.}
\label{fig:Fig2}
\end{figure*}

The eigenvalues $\lambda$ of $\rm{H_{\rm eff}}$ are solutions of the following characteristic equation, $\det (\rm{H_{\rm eff}}-\lambda \mathbb{I}_2)$. Solving this equation leads to the eigenvalues,
\begin{equation} \label{eq:eq21}
\lambda_{\pm}:=\frac{1}{2}(\Delta_m^1 + \Delta_m^2)  -\frac{{\rm i}}{4}(\rm{\Gamma^1_{\rm eff}}+\rm{\Gamma^2_{\rm eff}})  \pm \frac{\sigma}{4},
\end{equation}
where $\sigma:=\sqrt{16\tilde{J}_m^2+(2\delta{\Delta_m}-{\rm i}\delta\rm{\Gamma_{\rm eff}})^2}$ with $\delta{\Delta_m}:=\Delta^1_m - \Delta^2_m$ and $\delta\rm{\Gamma_{\rm eff}}:=\rm{\Gamma^1_{\rm eff}}-\rm{\Gamma^2_{\rm eff}}$. The eigenvalues of our system are given by \cref{eq:eq21} and displayed in \autoref{fig:Fig1}, where the eigenfrequencies ($\omega_{\pm}=\operatorname{Re}(\lambda_{\pm})$) and the dissipations ($\gamma_{\pm}=\operatorname{Im}(\lambda_{\pm})$) are shown in \autoref{fig:Fig1}(b,d) and \autoref{fig:Fig1}(c,e), respectively. In this figure, the parametric driving is off ($\chi=0$) and the solid (dashed) line depicts the case before (after) the perturbation of the system. In \autoref{fig:Fig1}(b,c) the perturbation induces a splitting of the $\rm{EP}$, while in \autoref{fig:Fig1}(d,e) it instead induces a shift of the $\rm{EP}$. In the former case, any mass deposition ($\delta m$) acts as a perturbation on the system, and induces a frequency shift ($\delta \omega$) through the relationship, $\delta m=\mathcal{R}^{-1}\delta \omega$ where $\mathcal{R}=\frac{\omega_m}{2 m}$ is the mass responsitivity and $m$ is the mass of the resonator supporting the deposition. Therefore, this mass deposition leads to a frequency shift that lifts up the $\rm{EP}$-degeneracy, resulting in a sensitivity enhancement at the $\rm{EP}$ (see double arrow in \autoref{fig:Fig1}(b,c)). In the latter case, the perturbation induces a shift in the dissipation instead, which can be seen as a dissipation mismatch ($\delta \gamma :=|\gamma_1-\gamma_2|$) between the mechanical resonators \cite{Wang_2019,Mao_2023}. This perturbation can also be thought of as induced by a mismatch between optical damping ($\delta \gamma :=|\Gamma_1-\Gamma_2|$), and may originate from optical fluctuations within the cavities. This scenario also leads to a sensing scheme as shown by the double arrow in \autoref{fig:Fig1}(d,e).  

A general observation from \autoref{fig:Fig1} is that after a splitting due to a mass deposition (\autoref{fig:Fig1}(b,c) ), the $\rm{EP}$  is destroyed and that limits the ability of the system to perform multiple sensing scenarii in a time sequence. Indeed, the enhanced performance is deteriorated after each sensing process since every next detection is conducted on the basis of being destroyed by the previous detection. In order to preserve $\rm{EP}$ in the parameter space after a perturbation, it is preferable to shift the $\rm{EP}$ instead of lifting it up. Such a sensor based on shifting the $\rm{EP}$ (see \autoref{fig:Fig1}(d,e) ) will therefore meet the condition of a non-demolition sensing mechanism. With a fine adjustment of gain and loss in the system, we expect a linewidth preservation of the eigenstates after optical perturbation in the sensing scheme based on shifting the $\rm{EP}$. This may improve the performance of such sensing scenario in the perspective of experiments related to the precision of dispersion measurement \cite{Mao_2023}. However, we stress that the sensitivity of this novel sensing scheme is somehow weak compared to the splitting scheme (compare the double arrows in \autoref{fig:Fig1} (b) and (d) . 

In order to enhance the sensitivity in the two aforementioned discussed sensing schemes, we turn on parametric driving ($\chi\neq0$). \autoref{fig:Fig2} displays the enhancement of the splitting when parametric driving is accounted for in the system. \autoref{fig:Fig2}(a)  and \autoref{fig:Fig2}(c) show the frequency splittings in the case of $\chi=0$ and $\chi=0.3\omega_m$, respectively. Their splitting ($\Delta\omega_{\pm}:=\omega_{\pm}^{\delta\omega}-\omega_{\pm}$) are displayed on \autoref{fig:Fig2}(b) and \autoref{fig:Fig2}(d), respectively. It can be seen that the resulting splitting at the $\rm{EP}$ is greater in \autoref{fig:Fig2}(d) than in \autoref{fig:Fig2}(b). This reveals the fact that parametric terms can be used to tune the performance of the sensor. Another observation is that, the parametric driving shifts the overall frequencies towards low frequencies (compare $y-{\rm axis}$ of \autoref{fig:Fig2}(a) and \autoref{fig:Fig2}(c). The same investigation can also be done in sensing based on shifting, and similar observations can be pointed out (not shown here). 

\begin{figure*}[tbh]
  \begin{center}
  \resizebox{1.0\textwidth}{!}{
  \includegraphics{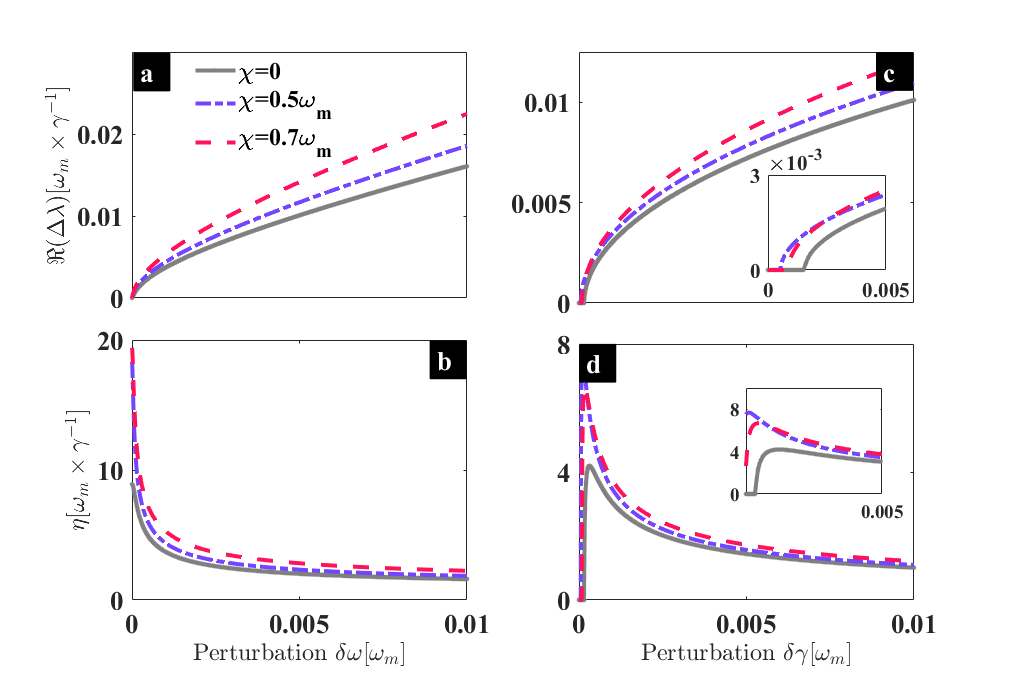}}
  \end{center}
  \caption{Performances of the sensing schemes. Panels (a,c): Sensitivities versus the perturbation for different values of $\chi$.  (b,d) Enhancement factors versus the perturbation for respective matching values of $\chi$ as in (a,c). Panels (a,b) are from the splitting sensing scheme, while panels (c,d) result from the sensing approach based on shifting the $\rm{EP}$. In the whole figure, full, dash-dotted and dashed lines correspond to $\chi=0$, $\chi=0.5\omega_m$, and $\chi=0.7\omega_m$, respectively. The rest of the parameters are the same as in \autoref{fig:Fig1}.}
  \label{fig:Fig3}
  \end{figure*}

In order to get more insight into the performance of these sensing schemes, we plotted in \autoref{fig:Fig3} the splitting and the enhancement factor. \autoref{fig:Fig3}(a,b) show these quantities for the sensing based on splitting while \autoref{fig:Fig3}(c,d) display the same quantities for the sensing based on shifting effect. The splitting represented here is the quantity ($\Delta\omega_{\pm}:=|\omega_{\pm}^{\epsilon}-\omega_{\pm}|$) at the $\rm{EP}$, while the enhancement factor is defined as being  this splitting over the perturbation  strength $\eta:=|\frac{\Delta\omega_{\pm}}{\delta\omega}|$. It is worth to keep in mind that the strength of the perturbation $\epsilon$ depend on the scenario, so that $\epsilon\equiv\delta\omega$ for the splitting case and $\epsilon\equiv\delta\gamma$ for the shifting one. In all these figures, the full line is without parametric driving ($\chi=0$), while the dashed-dotted and dashed lines are for $\chi:=0.5\omega_m$  and $\chi=0.7\omega_m$, respectively. It can be seen that parametric driving improves both the sensitivity and the enhancement factor. Furthermore, one can observe that these quantities scale faster in the splitting sensing scheme as compared to shifting scenario sensing (compare \autoref{fig:Fig3}(a) to \autoref{fig:Fig3}(c) or \autoref{fig:Fig3}(b) to \autoref{fig:Fig3}(d) ). One benefit of sensors operating under splitting mechanisms over those based on shifting is that the former performs very well for small perturbations, while the latter does not. Indeed, \autoref{fig:Fig3}(a,b) show giant enhancement sensitivity for small perturbations, and that is owing to the square root topological feature at the $\rm{EP}$. However, such an enhancement for small perturbations is missing in \autoref{fig:Fig3}(c,d), and is even zero (see inset). One reason behind this lack of sensitivity to small perturbations is that the shift is negligible for tiny disturbances. 
  
To understand the performance of these two sensing scenarii regarding the parametric term, one evaluates the quantity $\Delta\lambda_{\pm}=\lambda_{\pm}^{\epsilon}-\lambda_{\pm}$. This term is the difference between the perturbed eigenvalue $\lambda_{\pm}^{\epsilon}$ and the non-perturbed one. As stated above, our proposal can experience sensing through both splitting and shifting of the $\rm{EP}$. As we are interested in quantifying the sensitivity at the $\rm{EP}$, we can derive this difference of eigenvalues at the $\rm{EP}$ which yields (see \cref{App.B}),     
\begin{equation} \label{eq:eq22}
    \Delta\lambda_{\pm}^{\rm{EP}}:= \frac{\Delta \omega^{\epsilon}}{2}-\frac{{\rm i}|\delta\gamma|}{4}\pm \frac{\sigma_{\rm{EP}}^{\epsilon}}{2},
\end{equation}
where 
\begin{equation} \label{eq:eq23}
    \Delta \omega^{\epsilon}:= \frac{1}{2}\left[\frac{\omega_m\delta\omega}{\Delta_m^{2}}-\frac{1}{2}\left(\frac{\chi_1^{2}}{\omega_m}+\frac{\chi_2^{2}}{\omega_m}\right)\right],
\end{equation}
and 
\begin{equation} \label{eq:eq24}
    \sigma_{\rm{EP}}^{\epsilon}:= \sqrt{(8J_mr)^{2}+4\nu^2-4i\nu|\delta\gamma|+4\nu\mu-2i\mu|\delta\gamma|},
\end{equation}
with $\nu:=\frac{1}{2}\left(\frac{\chi_2^{2}}{\omega_m}-\frac{\chi_1^{2}}{\omega_m}\right)   
 - \frac{\omega_m\delta\omega}{\Delta_m^{2}}$ and $\mu:=-{\rm i}\delta\Gamma_{\rm{eff}}$. In these equations (\cref{eq:eq23} and \cref{eq:eq24}), all scenarii of perturbations allowed in our system have been considered. By considering that there is no parametric driving ($\chi_j=0$), and that there is no perturbation of the dissipation ($\delta\gamma=0$), the system reduces to the one investigated in \cite{Djorwe.2019}, and the expressions in \cref{eq:eq23}  and \cref{eq:eq24} reduce to $\Delta \omega^{\epsilon}= \frac{\delta\omega_2}{2}$ and $\sigma_{\rm{EP}}^{\epsilon}=(1+{\rm i})\sqrt{\frac{\delta\Gamma_{\rm{eff}}\delta\omega_2}{8}}$ as expected. This sensing scenario is based on a splitting at the $\rm{EP}$ as it can be seen in \autoref{fig:Fig1}(b,c). In that case, the amplification factor is given by:
 \begin{equation}\label{eq:eq25}
  \eta=\left|\frac{{\rm Re}{(\Delta\lambda_{\pm}^{\rm{EP}})}}{\delta\omega}\right|=\sqrt{\frac{\delta\Gamma_{\rm{eff}}}{8\delta\omega}}=\sqrt{\frac{m\delta\Gamma_{\rm{eff}}}{4\omega_m\delta m}}.   
 \end{equation}
When the perturbation is mainly from the dissipation ($\chi_j=0$, $\delta\omega=0$, $\delta\gamma\neq0$), which can be caused either by the mismatch damping of the two mechanical resonators or the fluctuation of the gain and loss induced by a driving fluctuation, both sensitivity and amplification factors are reduced to,
\begin{equation}\label{eq:eq26}
  \Delta\lambda_{\pm}^{\rm{EP}}:=-\frac{{\rm i}}{4}|\delta\gamma|\pm \frac{{\rm i}}{4}\sqrt{2|\delta\gamma|\delta\Gamma_{\rm{eff}}}
 \end{equation}
 and 
 \begin{equation}\label{eq:eq27}
  \eta=\left|\frac{{\rm Re}{(\Delta\lambda_{\pm}^{\rm{EP}})}}{\delta\gamma}\right|=\sqrt{\frac{\delta\Gamma_{\rm{eff}}}{8|\delta\gamma|}},   
 \end{equation}
which also scales as a square root of the perturbation. It should be highlighted that $\delta\gamma$ can be negative in order to make sense of the enhancement factor definition. Moreover, \cref{eq:eq26} shows that there is no effect of the perturbation on the reference frequency while there is a negative shift of $\frac{1}{4}|\delta\gamma|$ in the reference dissipation as seen in \autoref{fig:Fig1}(d,e). When parametric mechanical driving is involved ($\chi_j\neq0$), we expect to point out some improvement as depicted in \autoref{fig:Fig3}. By considering the sensing based on splitting scheme for instance ($\chi_j\neq0$, $\delta\omega\neq0$, $\delta\gamma \equiv 0$), one deduces from \cref{eq:eq22} the expression for the frequency difference at the $\rm{EP}$ as,     
 \begin{equation}\label{eq:eq28}
  \Delta\lambda_{\pm}^{\rm{EP}}:=\frac{1}{2}\left(\frac{\omega_m}{\Delta_m^2}\delta\omega-\frac{\chi^2}{\omega_m} \right) \pm \frac{1}{2}\sqrt{(8J_mr)^{2}+{\rm i}\frac{\omega_m}{\Delta_m^2}\delta\omega\delta\Gamma_{\rm{eff}}}.
 \end{equation}
In this expression, it results that the parametric driving shifts the reference frequency through $\frac{\chi^2}{\omega_m}$ as it appears in \autoref{fig:Fig2}(c) (compare $\rm{y-axis}$ of \autoref{fig:Fig2}(a) and \autoref{fig:Fig2}(c)), and also enhances the sensitivity. Indeed, the term ($8J_mr)^{2}$ which comes from the expansion of $\rm{\cosh(2r)}$ not only shifts the $\rm{EP}$ from the driving strength $\alpha^{\rm in}\sim 420 \omega_m^{1/2}$ (see 
\autoref{fig:Fig2}(a) ) to $\alpha^{\rm in}\sim500\omega_m^{1/2}$ (see \autoref{fig:Fig2}(c) ), but also contributes to the enhancement of the sensitivity as depicted in \autoref{fig:Fig3}(a, c). Furthermore, it should also be underlined that $\chi$ induces an amount of sensitivity improvement through the the term $\frac{\omega_m}{\Delta_m^2}$ under the square-root in \cref{eq:eq28}. Similar explanations regarding the effects of parametric driving on the sensitivity can be done for the sensing scenario based on the shifting mechanism. In fact, the expression of the frequency difference at the $\rm{EP}$ in this case is derived as,    
\begin{equation}\label{eq:eq29}
  \Delta\lambda_{\pm}^{\rm{EP}}=-\frac{\chi^2}{2\omega_m}-\frac{{\rm i}}{4}|\delta\gamma|\pm \frac{1}{4}\sqrt{(8J_mr)^{2}-2|\delta\gamma|\delta\Gamma_{\rm{eff}}}.  
 \end{equation}
From this expression, we can see that $\chi$ induces a shift of the reference frequency through $\frac{\chi^2}{2\omega_m}$ as seen while comparing the $\rm{y-axis}$ of \autoref{fig:Fig2}(a) to \autoref{fig:Fig2}(c). As in \autoref{fig:Fig1}(e), the shift in the dissipation is still captured by $\frac{|\delta\gamma|}{4}$. More importantly, in \cref{eq:eq29}, the sensitivity is mainly enhanced by the parametric driving through $(8J_mr)^{2}$. Therefore, these analytical expressions predict the improvement of both sensing schemes exhibited by our proposal, and are in good agreement with our findings. 

\subsection{Transmissions and output spectra}\label{Noise}
The current section analyzes the transmission and output spectra of the cavity fields. 
The effects of thermal noise, and the different involved perturbations will be investigated in order to figure out their impact on the EP feature. For that purpose, we start from the linearized dynamical equations of our system, and by moving to the frequency domain we derive the following scattering coefficients (see details in \cref{App.C}), 
\begin{subequations}\label{eq:Trans}
\small{
\begin{align}
 S_{11}(\omega)&:=1-\dfrac{2 \tilde{G}_1 \xi_2(\omega) \sqrt{\Gamma_1}}{\sqrt{\kappa} \xi_{\rm eff}(\omega) }, \,\,
 S_{12}(\omega) :=-\dfrac{2 {\rm i} \tilde{G}_1 \tilde{J}_m \sqrt{\Gamma_2}}{\sqrt{\kappa} \xi_{\rm eff}(\omega) },\\
 S_{13}(\omega) &:=\dfrac{2 {\rm i} \tilde{G}_1 \xi_2(\omega) \sqrt{\gamma_m}}{\sqrt{\kappa} \xi_{\rm eff}(\omega) } ,\,\,
 S_{14}(\omega) :=\dfrac{2 \tilde{G}_1 \tilde{J}_m \sqrt{\gamma_m}}{\sqrt{\kappa} \xi_{\rm eff}(\omega) },\\
 S_{21}(\omega)&:=\dfrac{2 {\rm i} \tilde{G}_2 \tilde{J}_m \sqrt{\Gamma_1}}{\sqrt{\kappa} \xi_{\rm eff}(\omega) } , \,\,
 S_{22}(\omega) :=1+\dfrac{2 \tilde{G}_2 \xi_1(\omega) \sqrt{\Gamma_2}}{\sqrt{\kappa} \xi_{\rm eff}(\omega) } ,\\
 S_{23}(\omega) &:=\dfrac{2 \tilde{G}_2 \tilde{J}_m \sqrt{\gamma_m}}{\sqrt{\kappa} \xi_{\rm eff}(\omega) },\,\,
 S_{24}(\omega) :=-\dfrac{2 {\rm i} \tilde{G}_2 \xi_1(\omega) \sqrt{\gamma_m}}{\sqrt{\kappa} \xi_{\rm eff}(\omega) }, 
 \end{align}   
 } 
 \end{subequations}
where  $\xi_{\rm eff}(\omega):= \xi_1(\omega) \xi_2(\omega) +\tilde{J}^2_m$ with  $\xi_1(\omega):= \Gamma^1_{\rm eff}/2 +{\rm i} (\Delta^{1}_m-\omega)$ and $\xi_2(\omega)=  \Gamma^2_{\rm eff}/2 +{\rm i} (\Delta^{2}_m-\omega)$. The coefficients $S_{11}$ and $S_{22}$ stand for the transmission of the first and second cavity, respectively.
\begin{figure*}[tbh]
\begin{center}
\resizebox{0.45\textwidth}{!}{
\includegraphics{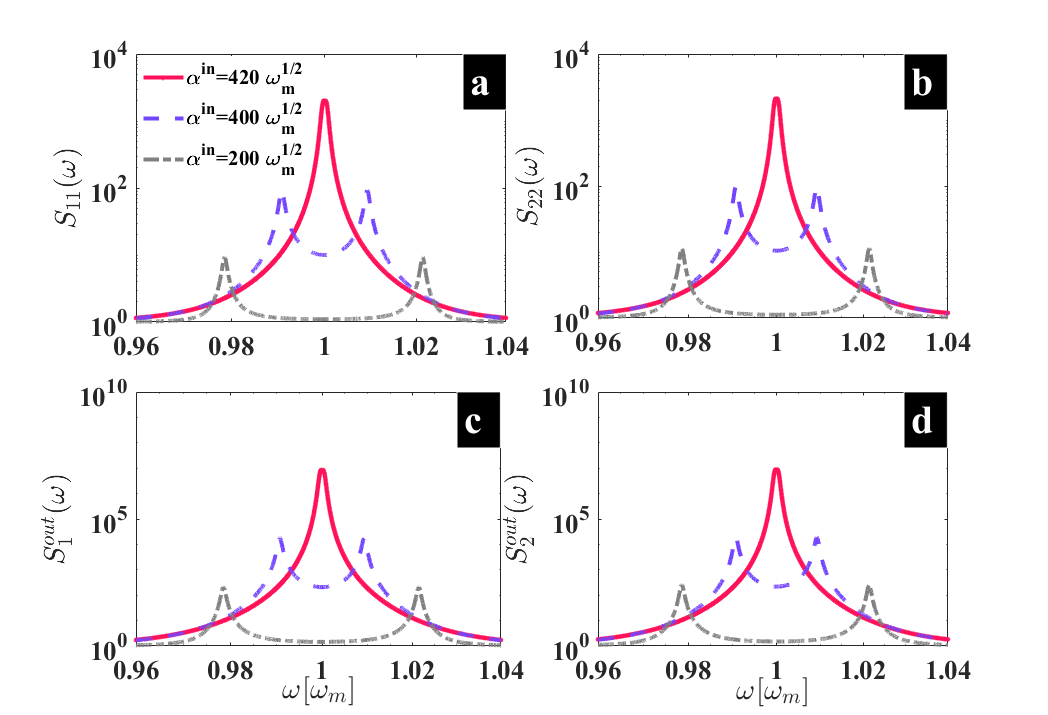}}
\resizebox{0.45\textwidth}{!}{
\includegraphics{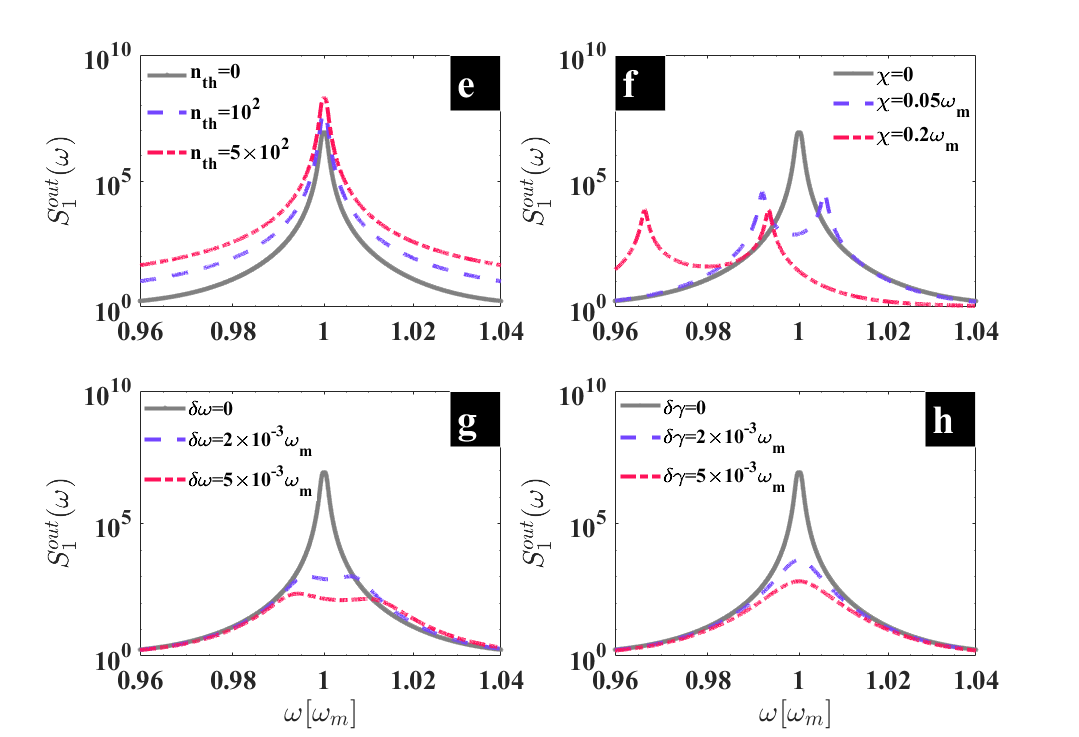}}
\end{center}
\caption{Output spectra. Panels (a,b): Transmission $S_{11}(\omega)$ and $S_{22}(\omega)$ of the first and second cavities, respectively. Panels (c,d) Output spectra $S_1^{\rm out}$ and $S_2^{\rm out}$  of the first and second cavities, respectively. The full line depicts the case at the EP for $\alpha^{\rm in}=420\omega_m^{1/2}$ (see \autoref{fig:Fig1} (b)). The dashed and dash-dotted lines correspond to cases away from the EP. Effect of the thermal noise in (e), and the parametric driving in (f), the frequency shift perturbation in (g), and dissipation mismatch perturbation in (h). The full lines captures the EP at $\alpha^{\rm in}=420\omega_m^{1/2}$ as in \autoref{fig:Fig1} (b), while the dashed and dash-dotted lines correspond to the perturbed cases depending on the varying parameter. The other parameters are the same as in \autoref{fig:Fig1}}
\label{fig:Fig4}
\end{figure*}
Owing to our assumption that the cavities are degenerated, these two coefficients become similar when the driving is strong enough (see full lines in \autoref{fig:Fig4}(a,b). For weak driving strength ($\alpha^{\rm in}\lesssim 90 \omega_m^{1/2}$), however, $S_{11}(\omega)$ shows two dips while $S_{22}(\omega)$ exhibits two peaks because they are far from the EP (not shown). This weak driving regime is interesting for nonreciprocal phenomena, and will be developed in our future investigations. The full lines in \autoref{fig:Fig4} (a,b) show the shape of transmissions, while in \autoref{fig:Fig4}(c,d) they depict output spectra at the EP. The output spectra can be expressed in terms of the given coefficients in \cref{eq:Trans} and yield (see \cref{App.C} for more detail ),
\begin{subequations}\label{eq:spec1}
\small{
\begin{align}
S^{(1)}_{\rm out}(\omega) &=\left| S_{11}(\omega )\right| {}^2+\left| S_{12}(\omega )\right| {}^2+\left(2N_{s_1}+1\right) \left| S_{13}(\omega )\right| {}^2\nonumber
\\&+\left(2N_{s_2}+1\right) \left| S_{14}(\omega )\right| {}^2,\\
S^{(2)}_{\rm out}(\omega) &=\left| S_{21}(\omega )\right| {}^2+\left| S_{22}(\omega )\right| {}^2+\left(2N_{s_1}+1\right) \left| S_{23}(\omega )\right| {}^2 \nonumber 
\\&+\left(2N_{s_2}+1\right) \left| S_{24}(\omega )\right| {}^2.
 \end{align}   
 } 
 \end{subequations}
where $N_{s_{1,2}}=(n_{th_{1,2}}+1) \sinh ^2(r_{1,2})+n_{th_{1,2}} \cosh ^2(r_{1,2})$, and $n_{th}$ the thermal population. As it can be seen, the shapes of these spectra are mainly depending on those of the transmission ones. The dashed and dash-dotted lines in \autoref{fig:Fig4}(a) to \autoref{fig:Fig4}(d) depict the situations away from the EP, and we can see two peaks which are reminiscent of  the eigenfrequencies shown in \autoref{fig:Fig1}(b) for instance. In what follows, we will only consider the spectrum $S_1^{\rm out}$ and will analyze the effects of some parameters on it as displayed from \autoref{fig:Fig4}(e) to \autoref{fig:Fig4}(h) . The effect of thermal noise $n_{th}$ on the output spectrum is shown in \autoref{fig:Fig4}(e), where we see that the EP is kept even though the output signal is amplified. This means that any output signal from our EP-sensor will be amplified, and this is due to the gain in our system. Such an amplification phenomenon is well-known in optomechanical systems, and seems to be beneficial for non-Hermitian sensing. Moreover, this thermal noise effect can be monitored by tuning the squeezing parameter $r$ through $N_{s}$. \autoref{fig:Fig4}(f) depicts the effect of parametric driving on the output spectrum, and it can be observed that $\chi$ shifts the spectrum's shape towards low-frequency values as expected (see \autoref{fig:Fig2}(c) for instance). Moreover, the driving strength at which EP happens has been increased. Owing to these changes in the input driving, the EPs corresponding to both the dashed and the dash-dotted lines are missing in \autoref{fig:Fig4}(f). To capture these EPs and to quantify how they have been shifted,  one needs to adjust the driving strength to $\alpha^{\rm in}\sim 130\omega_m^{1/2}$ for the dashed line and $\alpha^{\rm in}\sim 170\omega_m^{1/2}$ for the dash-dotted line. \autoref{fig:Fig4}(g) and \autoref{fig:Fig4}(h) display the effects of the perturbations on the frequency and the dissipation (or quality factor), respectively. The former corresponds to a sensing scheme based on the splitting of the EP, while the latter relies on sensing upon shifting of the EP. As aforementioned in \cref{Sen.EP}, it can be seen that the EP is destroyed in \autoref{fig:Fig4}(g) while it is preserved and shifted (down) in \autoref{fig:Fig4}(h). Moreover, the preserved peak of the EP in \autoref{fig:Fig4}(h) has been de-amplified and we stress that this may impair the performance of the sensing process.           

\section{Conclusion} \label{Con}
We investigated the effects of mechanical parametric driving on sensing performance at an exceptional point. This parametric behaviour comes from the modulation of the mechanical spring constant of the involved resonators. Two sensing schemes have been put forward. The sensing upon splitting of the EP as the frequency undergoes a perturbation, and the sensing through shifting of the EP that is induced by a damping mismatch for instance. It has been pointed out that mechanical parametric driving enhances the sensitivity of both schemes. Moreover, this parametric driving shifts the frequency at which EP happens towards low values. Furthermore, we figured out the fact that the sensitivity enhancement is more efficient under the splitting scenario than what is realized while shifting the EP. These results have also been confirmed through output spectra and transmission analysis of the optical cavities. These parametric effects on sensing reveal how nonlinearities can be used as requirement tools for better performances of sensors. This work provides a new way of improving EP-sensors and sheds light on the beneficial effects of nonlinear phenomena on non-Hermitian sensing.     

\section{Methods} \label{meth}
Our results are performed from numerical simulations and analytical calculations based on standard Quantum Langeving Equations (QLEs). These QLEs are derived from the system's Hamiltoninan following with their well known linearization process. From the linearized equations, an adiabatic elimination is carried out to trace out the intracavity flieds, leading to the effective mechanical Hamiltonian of the system. The real parts and the imaginary parts of the eigenvalues of this effective Hamiltonian capture the frequencies and the linewidths of the eigenmodes in the system. The coalescence of these frequencies and linewidths leads to an emergence of EP,  which is crucial to investigate sensor performances under perturbation and parametric effect. As the system is perturbed, the sensor sensitivity is captured from the split or the shift of the frequency and the linewidth of the EP. To refer on experimental scenario, the output spectra have been derived based on the standard output-input optomechanical relation. In practice, these spectra are used to externally infer any perturbation strength by monitoring the corresponding changes.
\section*{Acknowledgments}
\textbf{Funding:} This work has been carried out under the Iso-Lomso Fellowship at Stellenbosch Institute for Advanced Study (STIAS), Wallenberg Research Centre at Stellenbosch University, Stellenbosch 7600, South Africa. P. Djorwe acknowledges the receipt of a grant from the APS-EPS-FECS-ICTP Travel Award Fellowship Programme (ATAP), Trieste, Italy.
M. Asjad and D. Dutykh have been supported by the Khalifa University of Science and Technology under Award No. FSU- 2023-014.

\textbf{Author Contributions:} P.D. and M.A. conceptualized the
work and carry out the simulations and analysis. D.D., and B.D.-R. participated in all the discussions and provided useful suggestions to the final version of the manuscript. Y.P, D.D., and B.D.-R. supervised the work. All authors participated equally in the discussions and the preparation of the final manuscript.

\textbf{Competing Interests:} All authors declare no competing interests.

\section*{Data Availability}
Relevant data are included in the manuscript
and supporting information. Supplement data are available upon reasonable request.

\section*{Supplementary Materials}

\appendix \label{App} 

\section{Effective mechanical Hamiltonian} \label{App.A}
Some details about the derivation of  the effective mechanical Hamiltonian are presented in this Appendix. In the rotating wave approximation (RWA) and in the frame rotating at driving frequency $\omega_p^j + \omega_d$, where $\omega_p^j$ is the $j^{th}$ driving frequency, our system's Hamiltonian reads: 
\begin{subequations}\label{eq:eqa3}
\begin{align}   
H_{\rm{0M}} &= \sum_j \left(\tilde{\omega}_j b^{\dagger}_jb_j-\Delta_ja^{\dagger}_ja_j -g a^{\dagger}_ja_j (b^{\dagger}_j+b_j) \right),\\
H_{\rm{PA}}&= \frac{\chi}{2}\sum_j(b^{\dagger 2}_j+b^{2}_j),\\
H_{\rm{int}} &= -J_m(b_1b_2^{\dagger} + b_1^{\dagger}b_2),\\
H_{\rm{drive}} &= {\rm i} E \sum_j (a^{\dagger}_j - a_j),
\end{align}
\end{subequations}
where $\tilde{\omega}_j:=\omega_j - \omega_d$ and $\Delta_j:=\omega_p^j - \omega_c$.
This Hamiltonian can be diagonalized by introducing the squeezing transformation $S(r)=\rm{exp}[r(b^2-b^{\dagger  2})]$, where $\rm{r}$ will be figured out latter on. This operator transforms through the Bogoliubov transformation $b$ to $b_s$ as follows,
\begin{equation}\label{eq:eqa4}
 b_s:=b\cosh(r)+b^{\dagger}\sinh(r). 
\end{equation}
 From \cref{eq:eqa4}, it is straightforward to obtain, 
\begin{subequations}\label{eq:eqa5}
\begin{align}      
b_j&=b_{s_j} \cosh(r_j)- b^{\dagger}{s_j}\sinh(r_j)  \\
b^{\dagger}_j&=b{\dagger}_{s_j} \cosh(r_j)- b_{s_j}\sinh(r_j),
\end{align}
\end{subequations}
which can be used in \cref{eq:eqa3}. This leads to the following transformed Hamiltonian,
\begin{subequations}\label{eq:eqa6}
\begin{align}     
H_{\rm{0M}}&=\sum_j \left(\Delta_m^j b^{\dagger}_{s_j}b_{s_j}-\Delta_ja^{\dagger}_ja_j -\tilde{g} a^{\dagger}_ja_j (b^{\dagger}_{s_j}+b_{s_j})      \right), \\
H_{\rm{int}}&=-\tilde{J}_m(b_{s_1}b_{s_2}^{\dagger} + b_{s_1}^{\dagger}b_{s_2}),\\
H_{\rm{drive}}&={\rm i} E \sum_j (a^{\dagger}_j - a_j),\\
H_{\rm{diss}}&=-{\rm i}\frac{\kappa}{2}\sum_j a^{\dagger}_j a_j - {\rm i}\frac{\gamma_m}{2}\sum_j b^{\dagger}_{s_j}b_{s_j},
\end{align}
\end{subequations}
where the Hamiltonian $H_{\rm{diss}}$ captures the optical ($\kappa$) and the mechanical ($\gamma_1=\gamma_2=\gamma_m$) dissipations  in the system. Furthermore, the driving amplitude has been expressed as $E=\sqrt{\kappa}a^{\rm in}$, where $\alpha^{\rm in}$ is related to the input power $P_{\rm in}$ through $\alpha^{\rm in}=\sqrt{\frac{P_{\rm in}}{\hbar \omega_p}}$. The following parameters have been also defined,
\begin{subequations} \label{eq:eqa7}
    \begin{align}
    \tilde{g} &=g {\rm e}^{-r}, \\
    \Delta_m^j&=\sqrt{\tilde{\omega}_j^2-\chi^2},\\
    \tilde{J}_m&=J_m \cosh(2r).
\end{align}
\end{subequations}
The Quantum Langevin Equations (QLEs) of the Hamiltonian given in \cref{eq:eqa6} can be derived using the Heisenberg equation $\Dot{\mathcal{O}}={\rm i}[H,\mathcal{O}] +\mathcal{N}$, where $H$ is a Hamiltonian, $\mathcal{O}$ is an operator ($\mathcal{O}\in \{ b_{s_j}, a_j\}$, and  $\mathcal{N}$ is  the quantum noise related to the operator.  Following this procedure, the QLEs of our system yield, 
\begin{subequations}
\label{eq:eqa8}
\begin{align}
\Dot{a}_j &=-\left({\rm i} [\Delta_j + \tilde{g}(b_{s_j} + b^{\dagger}_{s_j})]+\frac{\kappa}{2}\right)a_j + \sqrt{\kappa}\alpha^{\rm in}_j  +\sqrt{\kappa} a^{\rm in}_j, \\
\Dot{b}_{s_j}&=-\left({\rm i}\Delta_m^j+\frac{\gamma_m}{2}\right)b_{s_j}+{\rm i}\tilde{J}_m b_{3-j}+{\rm i}\tilde{g}a^{\dagger}_ja_j+\sqrt{\gamma_m}\beta^{\rm in}_j. 
\end{align}
\end{subequations}
The nonlinear set of equations displayed in \cref{eq:eqa8} can be linearized by splitting the operators in terms of their mean values ($c-$numbers) and some amount of quantum fluctuations as $a_j=\alpha_j+\delta \alpha_j$ and $b_{s_j}=\beta_j+\delta \beta_j$, where $\alpha_j\equiv \langle a_j \rangle$ and $\beta_j\equiv \langle b_{s_j} \rangle$. This procedure leads to the classical mean field equations,
\begin{subequations} 
\begin{align}
    \Dot{\alpha_j}&=-\left({\rm i}\tilde{\Delta}_j + \frac{\kappa}{2}\right)\alpha_j+\sqrt{\kappa}\alpha^{\rm in}_j, \\
    \Dot{\beta_j} &=-\left({\rm i}\Delta_m^j+\frac{\gamma_m}{2}\right)\beta_j+{\rm i}\tilde{J}_m \beta_{3-j}+{\rm i}\tilde{g}|\alpha_j|^2
    \end{align}
\end{subequations}  
and to the fluctuations equations,
\begin{subequations} 
\label{eq:eqa9}
 \small{ 
\begin{align}
    \delta\Dot{\alpha}_j &=-\left({\rm i}\tilde{\Delta}_j +\frac{\kappa}{2}\right)\delta\alpha_j+ {\rm i}\tilde{G_j}(\delta\beta_j +\delta\beta^{\dagger}_j) +\sqrt{\kappa}\delta\alpha^{\rm in}_j , \\
       \rm{\delta\Dot{\beta}_j}&=-\left({\rm i}\Delta_m^j+\frac{\gamma_m}{2}\right)\delta\beta_j+{\rm i}\tilde{J}_m \delta\beta_{3-j}+{\rm i}\tilde{G_j}(\delta\alpha^{\dagger}_j+\delta\alpha_j) \nonumber \\
      &+\sqrt{\gamma_m}\delta{\beta}^{\rm in}_j,  
      \end{align}
      }
\end{subequations}
with $\tilde{\Delta}_j=\Delta_j+2\tilde{g} {\rm Re}(\beta_j)$, and $\tilde{G_j}=\tilde{g}\alpha_j$ where the intracavity field amplitude $\alpha_j$ is assumed to be real. In order to get an effective mechanical model describing \cref{eq:eqa9}, we need to trace out the optical equations. For that purpose, we introduce the following slowly varying operators with tildes, $\delta\alpha_j=\delta\tilde{\alpha_j}{\rm e}^{{\rm i}\tilde{\Delta}_j t}$, $\delta\alpha^{\rm in}=\delta\tilde{\alpha^{\rm in}}{\rm e}^{{\rm i}\tilde{\Delta}_j t}$, and $\delta\beta_j=\delta\tilde{\beta_j}{\rm e}^{-{\rm i}\Delta_m^j t}$. By using these slowly varying operators in \cref{eq:eqa9}, one gets
\begin{subequations} 
\label{eq:eqa10}
 \small{ 
\begin{align}    
    \Dot{\delta\tilde{\alpha_j}} &=-\frac{\kappa}{2}\delta\tilde{\alpha_j}+ {\rm i}\tilde{G_j}(\delta\tilde{\beta_j}{\rm e}^{-{\rm i}(\Delta_m^j+\tilde{\Delta}) t} +\delta\tilde{\beta_j}^{\dagger}{\rm e}^{{\rm i}(\Delta_m^j-\tilde{\Delta}_j) t}) \\&+\sqrt{\kappa}\delta\tilde{\alpha}^{\rm in}_j, \\
    \Dot{\delta\tilde{\beta_j}}&=-\frac{\gamma_m}{2}\delta\tilde{\beta_j}+{\rm i}\tilde{J}_m \delta\tilde{\beta}_{3-j}+{\rm i}\tilde{G_j}(\delta\tilde{\alpha_j}^{\dagger}{\rm e}^{{\rm i}(\Delta_m^j-\tilde{\Delta}_j) t}\\&+\delta\tilde{\alpha_j}{\rm e}^{{\rm i}(\Delta_m^j+\tilde{\Delta}_j) t})+\sqrt{\gamma_m}\delta\tilde{\beta}^{\rm in}_j.
    \end{align}
      }
\end{subequations}
As gain and losses are involved in our system, we assume that the first cavity is driven at the red sideband ($\tilde{\Delta}_1=-\Delta_m^j$) while the second one is driven at the blue sideband ($\tilde{\Delta}_2=\Delta_m^j$). Therefore, on the red sideband we get ($j=1$):
\begin{subequations} 
\label{eq:eqa11}
 \small{ 
 \begin{align}  
    \delta\tilde{\Dot{\alpha_j}} &=-\frac{\kappa}{2}\delta\tilde{\alpha_j}+ {\rm i}\tilde{G_j}(\delta\tilde{\beta_j} +\delta\tilde{\beta_j}^{\dagger}{\rm e}^{2i\Delta_m^j t}) +\sqrt{\kappa}\delta\tilde{\alpha^{\rm in}_j}, \\
  \delta\tilde{\Dot{\beta_j}}&=-\frac{\gamma_m}{2}\delta\tilde{\beta_j}+{\rm i}\tilde{J}_m \delta\tilde{\beta}_{3-j}+{\rm i}\tilde{G_j}(\delta\tilde{\alpha_j}^{\dagger}{\rm e}^{2 {\rm i }\tilde{\Delta}_j t}+\delta\tilde{\alpha_j})+\sqrt{\gamma_m}\delta\tilde{\beta}^{\rm in}_j,
    \end{align}
      }
\end{subequations}
while on the blue sideband we have ($j=2$):
\begin{subequations} 
\label{eq:eqa12}
 \small{ 
\begin{align} 
\delta\tilde{\Dot{\alpha_j}} &=-\frac{\kappa}{2}\delta\tilde{\alpha_j}+ {\rm i}\tilde{G_j}(\delta\tilde{\beta_j}{\rm e}^{-2i\Delta_m^j t} +\delta\tilde{\beta_j}^{\dagger}) +\sqrt{\kappa}\delta\tilde{\alpha}^{\rm in}_j, \\
    \delta\tilde{\Dot{\beta_j}}&=\rm{-\frac{\gamma_m}{2}\delta\tilde{\beta_j}+{\rm i}\tilde{J}_m \delta\tilde{\beta}_{3-j}+{\rm i}(\tilde{G_j}\delta\tilde{\alpha_j}^{\dagger}+\tilde{G_j}^{\ast}\delta\tilde{\alpha_j}{\rm e}^{2i\tilde{\Delta}_j t})+\sqrt{\gamma_m}\delta\tilde{\beta}^{\rm in}_j}.
   \end{align}
      }
\end{subequations}
In the RWA, the terms oscillating with high frequency can be ignored and the set of dynamical equations describing the system reduces to, 
\begin{subequations} 
\label{eq:eqa13}
 \small{ 
\begin{align}     
     \Dot{\delta\tilde{\alpha_1}} &=-\frac{\kappa}{2}\delta\tilde{\alpha_1}+ {\rm i}\tilde{G_1}\delta\tilde{\beta_1}  +\sqrt{\kappa}\delta\tilde{\alpha}^{\rm in}_1, \\
     \Dot{\delta\tilde{\alpha_2}} &=-\frac{\kappa}{2}\delta\tilde{\alpha_2}+ {\rm i}\tilde{G_2} \delta\tilde{\beta_2}^{\dagger} +\sqrt{\kappa}\delta\tilde{\alpha}^{\rm in}_2, \\     
    \Dot{\delta\tilde{\beta_1}}&=-\frac{\gamma_m}{2}\delta\tilde{\beta_1}+{\rm i}\tilde{J}_m \delta\tilde{\beta}_2+{\rm i}\tilde{G_1}^{\ast}\delta\tilde{\alpha_1}+\sqrt{\gamma_m}\delta\tilde\beta^{\rm in}_1,\\    
    \Dot{\delta\tilde{\beta_2}}&=-\frac{\gamma_m}{2}\delta\tilde{\beta_2}+{\rm i}\tilde{J}_m \delta\tilde{\beta}_1+{\rm i}\tilde{G_2}\delta\tilde{\alpha_2}^{\dagger}+\sqrt{\gamma_m}\delta\tilde\beta^{\rm in}_2.
  \end{align}
      }
\end{subequations}
In order to get the effective mechanical system, we need to trace out the intracavity field ($\delta \tilde{\alpha_j}$) in \cref{eq:eqa13}. For this purpose, let us introduce $\delta\tilde{\alpha_j}=\delta\alpha_m^j {\rm e}^{-\frac{\kappa}{2}t}$, which after using them in \cref{eq:eqa13} leads to,
\begin{subequations} 
\label{eq:eqa14}
 \small{ 
\begin{align}     
     \Dot{\delta\alpha_m^1} &=({\rm i}\tilde{G_1}\delta\tilde{\beta_1}  +\sqrt{\kappa}\delta\tilde{\alpha^{\rm in}_1}){\rm e}^{\frac{\kappa}{2}t}, \\
     \Dot{\delta\alpha_m^2} &=({\rm i}\tilde{G_2} \delta\tilde{\beta_2}^{\dagger} +\sqrt{\kappa}\delta\tilde{\alpha^{\rm in}_2}){\rm e}^{\frac{\kappa}{2}t}.
 \end{align}
      }
\end{subequations}
These equations can be integrated as follows
\begin{subequations} 
\label{eq:eqa15}
 \small{ 
\begin{align}       
\delta\alpha_m^1 &=\int_{-\infty}^t ({\rm i}\tilde{G_1}\delta\tilde{\beta_1}  +\sqrt{\kappa}\delta\tilde{\alpha^{\rm in}_1}){\rm e}^{\frac{\kappa}{2}\tau} d\tau, \\
\delta\alpha_m^2 &=\int_{-\infty}^t({\rm i}\tilde{G_2} \delta\tilde{\beta_2}^{\dagger} +\sqrt{\kappa}\delta\tilde{\alpha^{\rm in}_2}){\rm e}^{\frac{\kappa}{2}\tau}d\tau,
 \end{align}
      }
\end{subequations}
where in the weak coupling regime ($\kappa \gg \tilde{G_j}$), can be adiabatically integrated and yields:
\begin{subequations} 
\label{eq:eqa16}
 \small{ 
\begin{align}     
     \delta \tilde{\alpha}_1 &=\frac{2}{\kappa}({\rm i}\tilde{G_1}\delta\tilde{\beta_1}  +\sqrt{\kappa}\delta\tilde{\alpha}^{\rm in}_1), \\
     \delta \tilde{\alpha}_2 &=\frac{2}{\kappa}({\rm i}\tilde{G_2} \delta\tilde{\beta_2}^{\dagger} +\sqrt{\kappa}\delta\tilde{\alpha}^{\rm in}_2).
 \end{align}
      }
\end{subequations}
By using \cref{eq:eqa16} back in \cref{eq:eqa13}, we obtain the following effective mechanical system,
\begin{subequations} 
\label{eq:eqa17}
 \small{ 
\begin{align}      
    \Dot{\delta\beta_1}&=-({\rm i}\Delta_m^1+\frac{\rm{\Gamma_{\rm eff}}^1}{2})\delta\beta_1+{\rm i}\tilde{J}_m \delta\beta_2+{\rm i}\sqrt{\Gamma_1^{\ast}}\delta\alpha^{\rm in}+\sqrt{\gamma_m}\delta\beta^{\rm in}_1,\\
 \Dot{\delta\beta_2}&=-({\rm i}\Delta_m^2+\frac{\rm{\Gamma_{\rm eff}}^2}{2})\delta\beta_2+{\rm i}\tilde{J}_m \delta\beta_1+{\rm i}\sqrt{\Gamma_2}\delta\alpha^{in\dagger}+\sqrt{\gamma_m}\delta\beta^{\rm in}_2,
 \end{align}
      }
\end{subequations}
where optical induced damping is $ \rm{\Gamma}_j:=\frac{4|\tilde{G_j}|^2}{\kappa}$ and the effective damping are $\rm{\Gamma^{1,2}_{\rm eff}}=\gamma_m \pm  \rm{\Gamma}_{1,2}$. From now on, we will assume that $ \rm{\Gamma}_1= \rm{\Gamma}_2 \equiv \rm{\Gamma}$.  \cref{eq:eqa17} can be put in its compact form as
\begin{equation} \label{eq:eqa18}
\Dot{\mathcal{O}}=\mathcal{M}\mathcal{O}+{\rm i}\sqrt{\mathcal{K}} \mathcal{N}+\sqrt{\gamma_m}  \mathcal{O}^{\rm in}
\end{equation}
where $\mathcal{O}:= (\delta\beta_1, \delta\beta_2)^\top$, $\mathcal{K}:= (\Gamma_1^{\ast}, \Gamma_1)^\top$, $\mathcal{N}:= (\delta\alpha^{\rm in}, \delta\alpha^{in\dagger})^\top$
,$\mathcal{O}^{\rm in}:= (\delta\beta^{\rm in}_1, \delta\beta^{\rm in}_2)^\top$, and the matrix
\begin{equation} \label{eq:eqa19}
\mathcal{M}:=
\begin{bmatrix}
-\left({\rm i}\Delta^1_m+\frac{\rm{\Gamma^1_{\rm eff}}}{2}\right) & {\rm i}\tilde{J}_m \\
{\rm i}\tilde{J}_m & - \left({\rm i}\Delta_m^2+\frac{\rm{\Gamma^2_{\rm eff}}}{2}\right) 
\end{bmatrix}.
\end{equation}

\section{Splitting vs shifting sensing schemes} \label{App.B}
\begin{figure*}[tbh]
  \begin{center}
  \resizebox{0.48\textwidth}{!}{
  \includegraphics{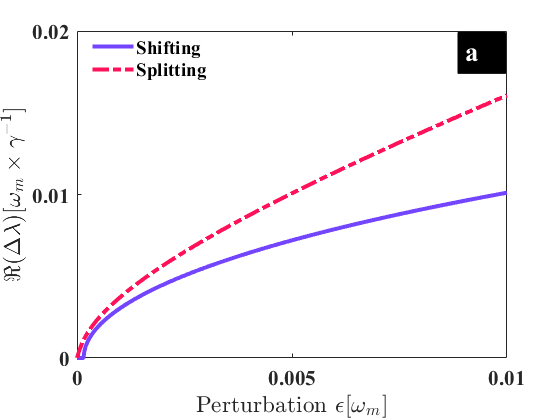}}
  \resizebox{0.48\textwidth}{!}{
  \includegraphics{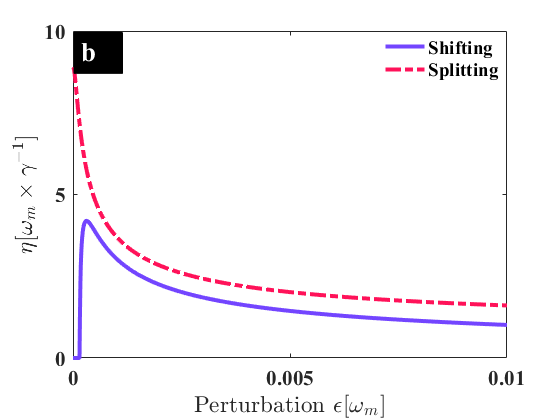}}
  \end{center}
  \caption{Sensing performances comparison. Sensitivities and the enhancement factors for the two sensing schemes when $\chi=0$.  The used parameters are the same as those in  \autoref{fig:Fig1}.}
  \label{fig:FigS1}
  \end{figure*}

This Appendix is meant to provide a comparison between the two sensing schemes investigated in the main text. We have shown the possibility of sensing based either the splitting or the shifting of the $\rm{EP}$, but a straight comparison between the two is missing. A general eigenvalue difference evaluated at the $\rm{EP}$ can be given by
\begin{equation} \label{eq:eqb1}
    \Delta\lambda_{\pm}^{\rm{EP}}= \frac{\Delta \omega^{\epsilon}}{2}-\frac{{\rm i}|\delta\gamma|}{4}\pm \frac{\sigma_{\rm{EP}}^{\epsilon}}{2},
\end{equation}
where 
\begin{equation} \label{eq:eqb2}
    \Delta \omega^{\epsilon}:= \frac{1}{2}\left[\frac{\tilde{\omega}_2\delta\omega_2}{\Delta_m^{2}}-\frac{\chi_2\delta\chi_2}{\Delta_m^{2}}-\frac{\chi_1\delta\chi_1}{\Delta_m^{1}}-\frac{1}{2}\left(\frac{\chi_1^{2}}{\tilde{\omega}_1}+\frac{\chi_2^{2}}{\tilde{\omega}_2}\right)\right]
\end{equation}
and 
\begin{equation} \label{eq:eqb3}
\sigma_{\rm{EP}}^{\epsilon}:= \sqrt{(8J_mr)^{2}+4\nu^2-4i\nu|\delta\gamma|+4\nu\mu-2i\mu|\delta\gamma|},
\end{equation}
with $\nu:=\frac{1}{2}\left(\frac{\chi_2^{2}}{\tilde{\omega}_2}-\frac{\chi_1^{2}}{\tilde{\omega}_1}\right)   
 +\frac{\chi_2\delta\chi_2}{\Delta_m^{2}}- \frac{\tilde{\omega}_2\delta\omega_2}{\Delta_m^{2}}-\frac{\chi_1\delta\chi_1}{\Delta_m^{1}}$ and $\mu:=(2\delta\tilde{\omega}-{\rm i}\delta\Gamma_{\rm{eff}})$. In these expressions, $\delta\omega_2$ and $\delta\gamma$ are the two types of perturbation used in the main text. The former is induced by a mass deposition on the second mechanical resonator, while the latter comes from optical fluctuations or a mismatch damping of the two mechanical resonators.  We stress that there can be also a perturbation induced by the fluctuation of the parametric driving which can be captured through $\delta\chi_j$ in \cref{eq:eqb3}.  However, we have not considered such a perturbation in our investigation in the main text. We have also made some assumptions for a seek of simplicity, $\tilde{\omega}_1=\tilde{\omega}_2\equiv\omega_m$ and $\chi_1=\chi_2 \equiv \chi$. With these approximations, we were able to derive different expressions for sensitivity and enhancement factor presented in the main text, which were qualitatively in good agreement with our results. In order to give a direct comparison between the two sensing scenarii, we plotted both their sensitivity and enhancement factor in \autoref{fig:FigS1}. The full line is from the sensing based on the shifting mechanism while the dashed line refers to splitting at the $\rm{EP}$. It can be clearly seen that the scheme based on the splitting performs better as discussed in the main text.
  
\section{Transmission and  output spectra}\label{App.C}
The measurement of cavity field fluctuations is commonly performed more conveniently in the frequency domain rather than the time domain experimentally. Hence, employing the Fourier transform definition for an operator $\mathcal{O}(\omega) = \int^{\infty}_{-\infty}   \mathcal{O}(t) {\rm e}^{-{\rm i}\omega t} dt$, Eq.(\ref{eq:eqa18}) can be written in Fourier domain as 
\begin{equation}\label{eq:eqc1}
 \mathcal{O}(\omega)= \mathcal{D}(\omega) \left[ {\rm i}\sqrt{\mathcal{K}} \mathcal{N}(\omega)+\sqrt{\gamma_m} \mathcal{O}^{\rm in}(\omega)\right],
\end{equation}
where $\mathcal{D} (\omega):=(- \mathcal{M}-{\rm i}\omega)^{-1}$.  We are interested to obtain the spectra  two optical modes at the output of the opto-mechanical cavity. According to the input-output theory, the operators for the output fields are related to the cavity and to the input noise operators by the relation $\delta \alpha^{\rm out}_1(\omega):= \sqrt{\kappa} \delta \alpha_1(\omega)-\delta \alpha^{\rm in}_1(\omega)$ and $\delta \alpha^{\rm out}_2(\omega):= \sqrt{\kappa} \delta \alpha_2(\omega)-\delta \alpha^{\rm in}_2(\omega)$. Then by using Eq.(\ref{eq:eqa16}) and Eq.(\ref{eq:eqc1}), the expressions for the two transmitted fields in Fourier domain are given by, 
\begin{subequations}
\label{eq:out}
\small{
 \begin{align}
\delta \alpha^{\rm out}_1(\omega)&= S_{11}(\omega) \delta  \alpha^{\rm in}_1(\omega) + S_{12}(\omega) \delta  \alpha^{in^\dagger}_2(\omega) + S_{13}(\omega) \delta \beta^{\rm in}_1(\omega) \nonumber 
\\ 
&+  S_{14}(\omega) \beta^{\rm in}_2(\omega),\\
\alpha^{out^\dagger}_2(\omega) &= S_{21}(\omega) \delta  \alpha^{\rm in}_1(\omega) + S_{22}(\omega) \delta  \alpha^{in^\dagger}_2(\omega)+ S_{23}(\omega) \delta \beta^{\rm in}_1(\omega)  \nonumber 
\\ & +  S_{24}(\omega) \beta^{\rm in}_2(\omega)
\end{align}
}
\end{subequations}
where 
\begin{subequations}
\label{eq:trans}
\small{
\begin{align}
 S_{11}(\omega):=1-\dfrac{2 \tilde{G}_1 \xi_2(\omega) \sqrt{\Gamma_1}}{\sqrt{\kappa} \xi_{\rm eff}(\omega) }, 
 \,\,
 S_{12}(\omega) :=-\dfrac{2 {\rm i} \tilde{G}_1 \tilde{J}_m \sqrt{\Gamma_2}}{\sqrt{\kappa} \xi_{\rm eff}(\omega) },\\
 S_{13}(\omega) :=\dfrac{2 {\rm i} \tilde{G}_1 \xi_2(\omega) \sqrt{\gamma_m}}{\sqrt{\kappa} \xi_{\rm eff}(\omega) } ,
 \,\,
 S_{14}(\omega) :=\dfrac{2 \tilde{G}_1 \tilde{J}_m \sqrt{\gamma_m}}{\sqrt{\kappa} \xi_{\rm eff}(\omega) },\\
 S_{21}(\omega)=\dfrac{2 {\rm i} \tilde{G}_2 \tilde{J}_m \sqrt{\Gamma_1}}{\sqrt{\kappa} \xi_{\rm eff}(\omega) } , 
 \,\,
 S_{22}(\omega) :=1+\dfrac{2 \tilde{G}_2 \xi_1(\omega) \sqrt{\Gamma_2}}{\sqrt{\kappa} \xi_{\rm eff}(\omega) } ,\\
 S_{23}(\omega) :=\dfrac{2 \tilde{G}_2 \tilde{J}_m \sqrt{\gamma_m}}{\sqrt{\kappa} \xi_{\rm eff}(\omega) },
 \,\,
 S_{24}(\omega) :=-\dfrac{2 {\rm i} \tilde{G}_2 \xi_1(\omega) \sqrt{\gamma_m}}{\sqrt{\kappa} \xi_{\rm eff}(\omega) } ,
 \end{align}
      }
\end{subequations}
where  $\xi_{\rm eff}(\omega)= \xi_1(\omega) \xi_2(\omega) +\tilde{J}^2_m$ with  $\xi_1(\omega)= \Gamma^1_{\rm eff}/2 +{\rm i} (\Delta^{1}_m-\omega)$ and $\xi_2(\omega)=  \Gamma^2_{\rm eff}/2 +{\rm i} (\Delta^{2}_m-\omega)$. The  input quantum noise operators cavity and mechanical modes are characterized by the following correlations
\begin{subequations} 
\label{eq:Suscep}
\begin{align}
\langle \delta \alpha^{\rm in}_j(t), \delta \alpha^{in^\dagger}_{j}(t')\rangle &= \delta(t+t'),  \\
\langle \delta \beta^{in^\dagger}_{j}(t'), \delta \beta^{\rm in}_j(t)\rangle &= N_s  \delta(t+t'),\\
\langle \delta \beta^{\rm in}_j(t), \delta \beta^{\rm in}_{j}(t')\rangle & =  M_s \delta(t+t'), \quad   { \rm for} \,\, \, \,  j \in \{ 1, 2 \}, 
\end{align}
\end{subequations}
where $N^j_s= (n^j_m+1) \sinh ^2(r_j)+n^j_m\cosh ^2(r)$ and $M^j_s=(2 n^j_m+1) \sinh (r_j) \cosh (r_j)$ with $n^j_m=[{\rm e}^{\hbar \omega_j/K_B T}-1]^{-1}$ is the  mean thermal occupation number of the mechanical modes. Here we assumed that both mechanical resonators are kept at same temperature $T$. By using the expressions from \cref{eq:out} and \cref{eq:Suscep}, one can derive the output spectra defined as
\begin{subequations}\label{eq:spec}
\begin{align}
S^{(1)}_{\rm out}(\omega) &=\int [\langle a^{out \dagger}_1(\omega)\,\, a^{\rm out}_1(\omega') \rangle+\langle a^{\rm out}_1(\omega)\,\, a^{out\dagger}_1(\omega') \rangle] d\omega', \nonumber \\
&=\left| S_{11}(\omega )\right| {}^2+\left| S_{12}(\omega )\right| {}^2+\left(2N_{s_1}+1\right) \left| S_{13}(\omega )\right| {}^2 \nonumber 
\\& + \left(2N_{s_2}+1\right) \left| S_{14}(\omega )\right| {}^2,\\
S^{(2)}_{\rm out}(\omega) &=\int [\langle a^{out \dagger}_2(\omega)\,\, a^{\rm out}_2(\omega') \rangle+\langle a^{\rm out}_2(\omega)\,\, a^{out\dagger}_2(\omega') \rangle] d\omega', \nonumber \\
&=\left| S_{21}(\omega )\right| {}^2+\left| S_{22}(\omega )\right| {}^2+\left(2N_{s_1}+1\right) \left| S_{23}(\omega )\right| {}^2 \nonumber
\\&+\left(2N_{s_2}+1\right) \left| S_{24}(\omega )\right| {}^2,
\end{align}
\end{subequations}
where $N_{s_{1,2}}:=(n_{m_{1,2}}+1) \sinh ^2(r_{1,2})+n_{m_{1,2}} \cosh ^2(r_{1,2})$.

\bibliography{Sensor}

\end{document}